# Dynamical Jumping Real-Time Fault-Tolerant Routing Protocol for Wireless Sensor Networks

Guowei Wu [1], Chi Lin [1], Feng Xia [1,*], Lin Yao [1,2,*], He Zhang [1] and Bing Liu [1]

[1] School of Software, Dalian University of Technology, Dalian 116620, China;
   E-Mails: wgwdut@dlut.edu.cn (G.W.); linchix@gmail.com (C.L.);
   fofospeak@live.cn (H.Z.); liubing-051@live.cn (B.L.)
[2] School of Electronics & Information, Dalian University of Technology, Dalian 116021, China

* Author to whom correspondence should be addressed; E-Mails: f.xia@ieee.org (F.X.); yaolin@dlut.edu.cn (L.Y.)

**Abstract:** In time-critical wireless sensor network (WSN) applications, a high degree of reliability is commonly required. A dynamical jumping real-time fault-tolerant routing protocol (DMRF) is proposed in this paper. Each node utilizes the remaining transmission time of the data packets and the state of the forwarding candidate node set to dynamically choose the next hop. Once node failure, network congestion or void region occurs, the transmission mode will switch to jumping transmission mode, which can reduce the transmission time delay, guaranteeing the data packets to be sent to the destination node within the specified time limit. By using feedback mechanism, each node dynamically adjusts the jumping probabilities to increase the ratio of successful transmission. Simulation results show that DMRF can not only efficiently reduce the effects of failure nodes, congestion and void region, but also yield higher ratio of successful transmission, smaller transmission delay and reduced number of control packets.

**Keywords:** real-time; fault-tolerance; wireless sensor networks; routing protocol

## 1. Introduction

Wireless sensor networks (WSNs) have very broad application prospects in military disaster monitoring, environmental and ecological monitoring, earthquake, fire emergencies, medical systems, urban transportation and security monitoring [1]. In most of the applications, especially in emergent situations, real-time and fault tolerance characteristics are highly required. Nodes in WSN are prone to failure due to energy depletion, hardware failure, communication link errors, malicious attack, *etc*. [2]. Fault-tolerance is the ability of a system to deliver a desired level of functionality in the presence of faults. Extensive work has been done on fault tolerance and it has been one of the most important research topics in WSNs. Fault tolerance in a WSN system may exist at hardware layer, software layer, network communication layer, and application layer [2]. In this paper, we focus on the fault tolerance in real-time routing protocol level. To guarantee the real time performance of the nodes, each data packet is constrained in a time interval in which it must be sent to the destination node. If time expires, the data packet has to be discarded. Once node failure or congestion occurs, large amounts of data packets will be discarded, which may cause disastrous consequences. Consequently, it is more significant and challenging to provide both real-time and fault tolerance characteristics in WSN routing protocol.

Existing real-time fault tolerant WSN routing protocols adopt hop-by-hop transmission mode [3-5], where the next hop node is selected based on the transmission time estimation or the relations of the distance among the current node, neighbor node and sink node. Failure nodes are treated as an empty area (VOID), and data packets are sent to the sink node *via* bypass. However, these methods do not predict network congestion in advance [1], and the remaining transmission time of the data packet is only used for checking the validity of the data packets. When a data packet cannot be transmitted to the next hop node, it will be automatically discarded at once, which wastes the transmission energy. Moreover, the upper stream node cannot receive the feedback information from the current node and thus affect the subsequent transmission.

In this paper, a dynamical jumping real-time fault-tolerant routing protocol is proposed, namely DMRF. Each node utilizes the remaining transmission time of the data packets and the state of the forwarding candidate node set to dynamically select the next hop. Once node failure, network congestion or void region occurs, the transmission mode will switch to jumping mode, which aims at reducing the transmission time delay and ensuring the data packets to be sent to the destination node within the specified time limit. According to the feedback information from the downstream node, each node dynamically adjusts the jumping probabilities to increase the ratio of successful transmission.

The remainder of this paper is organized as follows. Section 2 introduces related work. Section 3 provides the application scenario of DMRF protocol. Section 4 presents the DMRF protocol in detail. Section 5 provides the feasibility proof and performance analysis of DMRF. The performance of DMRF is evaluated in Section 6. Section 7 concludes the paper and outlines some future work.

## 2. Related Work

The growing interest in WSNs and the continual emergence of new techniques has inspired some efforts to design real-time fault tolerant routing protocols in this area. In [6], a distributed transmission scheduling strategy was adopted. Non-conflict messages are used to achieve real-time scheduling.

A time allocation scheduling method was designed to prevent the communication channel collision. A dynamic rate adjustment routing protocol based on energy consumption, *i.e.*, RPAR (Real-time Power-Aware Routing), was proposed in [7], in which the node transmitting power is dynamically adjusted according to its energy consumption. However, RPAR doesn't consider the effect of node failure on real-time transmission. In [8], an energy-aware QoS routing protocol was proposed, which can find the transmission path that both has the maximum energy utilization and can meet the end-to-end transmission requirement. However, the time delay caused by channel is ignored. Moreover, the time complexity is high, which makes the method not suitable for the computation resource-constrained environment. In [9], the authors proposed a highly energy-efficient real-time routing protocol, which uses the constraint equivalent delay (CED) to dynamically select the next hop, and can simplify the process of finding the routing path greatly. This protocol can not only reduce the energy cost in the real time transmitting period, but also reduce the delay of end-to-end transmission. However, the node failure and network congestion are not considered during the process of real-time transmission. A cluster-based real-time transmission scheme was proposed in [10]. It performs data fusion in appropriate time, which can reduce energy consumption and the time of queuing in the buffer. However, frequent data fusion may cause higher transmission delay, and thus affect the real-time performance. SPEED (Stateless Protocol for Real-Time Communication in Sensor Networks) is one of the most classical real-time routing protocols for WSNs [3], which estimates the transmission rate between the current node and the sink node and establishes the transmission path according to the estimated rate to ensure the data can be sent to the sink node before the deadline. The SPEED protocol doesn't consider the effect of node failure and congestion. When node failure and network congestion occur, the information cannot be fed back to upstream nodes [11] timely, which thus affects the subsequent transmission and causes the relevant packet to be discarded. In [3], two heuristic methods called SPEED-T and SPEED-S were also proposed, which select the next hop in the Forwarding Candidate Set (FCS) according to the minimal transmission delay and the fastest transmission rate, respectively. SPEED-T and SPEED-S enhanced the real-time performance of SPEED, but they do not overcome the drawback of SPEED. In [12], a two-hop based real-time routing protocol was developed, which selects the next hop by evaluating the node in the two-hop range. However, it cannot overcome the drawbacks of SPEED either. A multi-path and multi-level SPEED routing protocol (MMSPEED) was proposed in [4], which dynamically selects the next hop according to the distance among the current node, neighbor node and sink node and sets up a tree structure for transmission according to the reachability of nodes. However, the time complexity of estimating the reachability of the nodes is an exponential function of the distance between the current node and the sink node. Therefore, it is not suitable for large-scale long-distance transmission. Furthermore, if congestion occurs in the initial hop nodes, the global transmission will be affected greatly. A real-time fault tolerant routing protocol called FTSPEED was proposed in [5]. FTSPEED is based on SPEED. It provides the transmission path selection method in the case that the next hop of current node is void. The data can be sent to the sink *via* bypassing the empty areas. FTSPEED reduces the impact of the void region, but the transmission path length maybe considerably long, which may ultimately cause the data packet failed to be sent to the sink node before deadline. Energy cost for setting up the routing path has been studied in [7-11,13-15].

Although the existing schemes [1-14,16-27] play important roles in improving WSN performance, real-time fault tolerant routing protocol design is still a challenging area in WSNs. In this paper, a dynamic jumping real-time fault tolerant routing protocol based on the research of other relevant protocols is demonstrated. The major difference between this work and the aforementioned protocols includes the following aspects:

(1) In order to guarantee real-time and fault-tolerant characteristics, jumping transmission mode is adopted. When node failure, network congestion or empty region is detected, or the remaining transmission time of the data packet is near to deadline, jumping transmission mode will be used to reduce the transmission delay, thus ensuring the data packets are sent to the destination node in specified time limit.

(2) Feedback mechanism is exploited to enhance the successful transmission ratio. Each node feeds back the information about node failure, network congestion and empty area to its upstream node and the message is forwarded to the data sources. Then the jumping probability can be dynamically adjusted by using the feedback information, which can prevent the subsequent transmission from the effect caused by failure node, network congestion or empty region.

(3) When using the hop-by-hop transmission mode, the node in FCS with the minimum times of transmission is selected as the next hop node, through which the average energy cost of each node can be balanced. Therefore, the whole network life time can be prolonged.

## 3. Preliminaries

In this section, we describe the application scenario of the DMRF protocol. Various variables used in the DMRF protocol will be defined. Here, we consider a scenario where the WSN is formed by stationary sensors in a two-dimension sensing field. Figure 1 shows the scenario. The source node collects relevant data, and transmits through intermediate nodes to the sink node. The nodes deployed in the network conform to standards or random distribution. The initial energy of each node is homogeneous and the batteries are not rechargeable. The communication radius of the node is $r$. In the network topology, there exist failure nodes, empty area and congestion area. Data packets are required to be sent from the source node to the sink node in the specified time limit. For each node, any other node in its communication range could become its neighbor. The DMRF protocol aims to reduce the effect of failure nodes, empty area and congestion area in the network to guarantee real-time performance. We first give the definitions of some variables used in the DMRF protocol, as listed in Table 1.

**Table 1.** Definitions of variables and notations used in the DMRF protocol.

| Variable | Description |
| --- | --- |
| FAULTY & JFAULTY | Faulty & JFaulty state |
| CONG & JCONG & NORMAL | Congestion & JCongestion & Normal state |
| VOID | Void region |
| $Suc_j$ | Successful transmission ratio to node $j$ |
| $p_i$ | Jumping probabilities to node $i$ |
| $\lambda$ | Remaining transmission time factor |

**Table 1.** *Cont.*

| | |
|---|---|
| $\theta_{low}$, $\theta_{high}$ | Lower or Upper threshold of remaining transmission time factor |
| $\theta_{jump}$ | Jumping threshold |
| $delay_{i,j}$ | Delay between node *i* and node *j* |
| LOW & MEDIUM & HIGH | LOW & MEDIUM & HIGH transmission rate of data packets |
| T | Estimated transmission time |
| L | Remaining transmission time of data packet |
| v | Average transmission rate |
| t | Maximum remaining transmission time of data packet |
| h | Average length of one hop |
| e | Average energy consumption of each transmission |
| d | Node density |
| r | Sensing radius of node |
| c | Confidence variable of node |

**Figure 1.** System scenario.

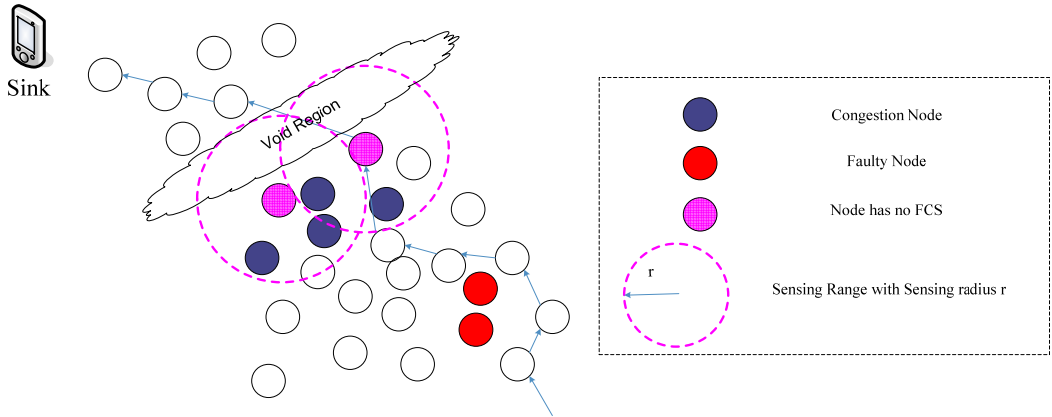

**Definition 1:** Node Failure State **FAULTY**. **FAULTY** indicates that the node is in failure state. If the states of the **FCS** nodes of the current node are all **FAULTY**, then the state of current node is set to **JFAULTY**, indicating the jumping transmission mode should be used to transmit data packet due to node failure.

**Definition 2:** Node Congestion State **CONG**. **CONG** indicates that the node is in congestion state. If the states of the **FCS** nodes of the current node are all **CONG**, then the state of current node is set to **JCONG**, indicating the jumping transmission mode should be used to transmit data packet due to the congestion.

**Definition 3:** Empty Area State **VOID**. **VOID** indicates that there is no node in the area. If the states of the **FCS** nodes of the current node are all **VOID**, then the state of the current node is set to **VOID**, indicating the jumping transmission mode should be used to transmit data packet due to the empty area.

**Definition 4:** Jumping Probability $p_i$. The $p_i$ indicates the probability of jumping from current node to the node *i*, the higher value of $p_i$ indicates the higher jumping probability.

**Definition 5:** Confidence variable *c*. The *c* indicates the confidence level of reliability of the node, the higher value of *c* indicates the higher reliability of the node.

## 4. DMRF Protocol

The transmission process with the DMRF protocol is depicted in Figure 2. The process is divided into five stages: initialization, data transmission, jumping transmission, jumping probability adjustment and transmission finish.

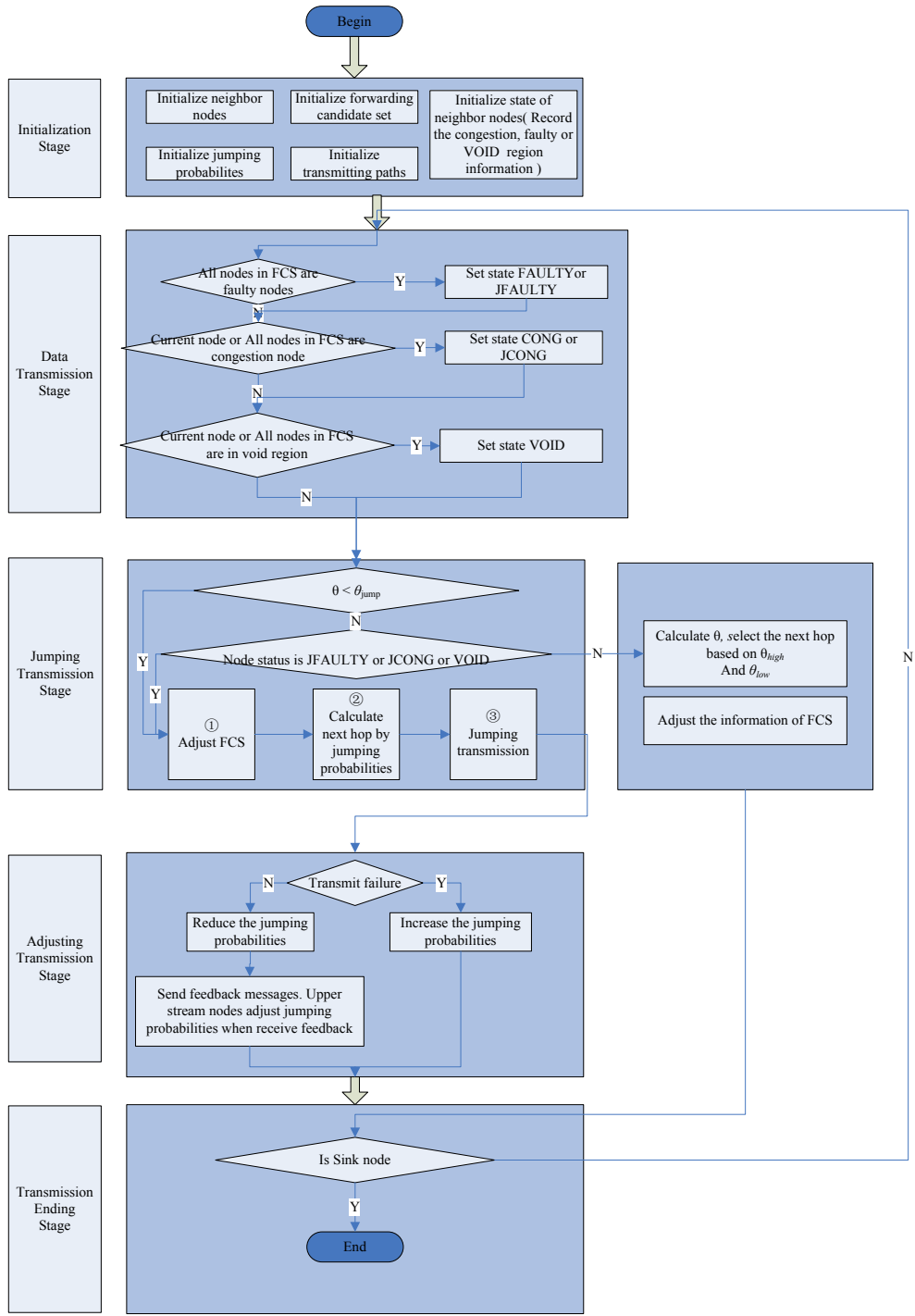

**Figure 2.** Data transmission with DMRF.

Each node maintains a dynamic routing table which is updated timely to record the states of the surrounding nodes. In initialization phase, DMRF initializes the neighbor nodes list, state list (records the information of node congestion, failure and empty area), **FCS** list, jumping probability table, and the initial transmission path. In data transmission phase, DMRF detects node failure, network congestion and VOID region, the detection methods will be described in subsections 4.1, 4.2 and 4.3, respectively. The left time of data packet will be checked to analyze if that packet need to be transmitted in a jumping mode. If none of the above conditions happen, DMRF will dynamically select a member from FCS as the next hop based on the transmission rate of data packet and local information. Once failure nodes, congestion nodes or VOID area are detected, or the left time factor of the data packet is less than the jumping threshold, the jumping transmission mode will be used. In the jumping transmission phase, each node dynamically adjusts the content of **FCS** (e.g., DMRF updates the member status in **FCS**). Then DMRF calculates the next hop node according to the jumping probability. By utilizing the jumping transmission mode, the data packet can jump over failure nodes, congestion nodes or VOID region, but it cannot guarantee the success of the jumping transmission, therefore the jumping probability adjustment phase is performed after each jumping transmission. In the jumping probability adjustment phase, DMRF adjusts the jumping probability according to the result of jumping transmission (success or failure) and feeds back the information to its upstream node (e.g., If node *a* fails to send a message to node *b*, *a* will immediately adjust its jumping probabilities, also it will feed back this information to its upstream node not to send messages to node *b*). When the data packet is sent to the sink node, the transmission is finished. The design principle of DMRF is described below.

DMRF adopts the similar mechanism used in [28], in which node can directly transmit data to sink node. The node selects the next hop node according to the jumping probability. If data transmission fails, the jumping probability is adjusted via feedback mechanism, which can not only avoid the effect caused by node failure, network congestion and VOID area, but also improve the transmission rate and make the possibly discarded data packet continue to be transmitted and thus lower the energy consumption of retransmission. The jumping transmission method is shown in Figure 3.

**Figure 3.** Jumping transmission.

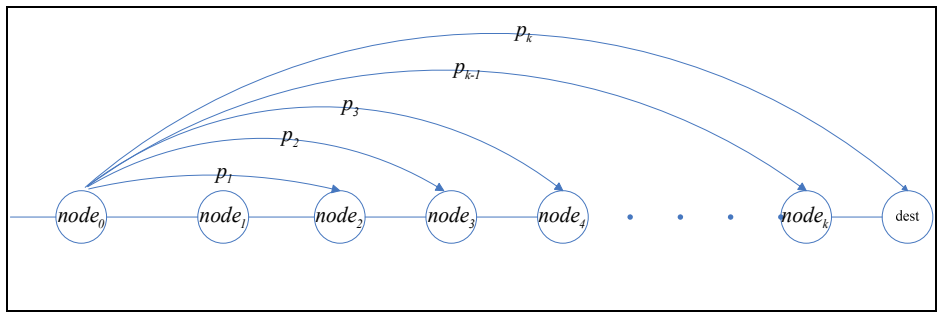

*4.1. Faulty Node Detection Method*

In this paper, the faulty node is defined as the node that cannot collect data or transmit data. **Algorithm 1** is the faulty node detection and node state adjustment algorithm. It checks the state of **FCS** to determine whether the node is faulty. A message is sent to the **FCS** nodes. According to the reply, it determines the node state and then updates the content of **FCS** nodes (state, transmission rate and delay). We utilize a variable *c* to represent the confidence of a node in **FCS**. Originally the value of each confidence variable is set as 100. Once transmission fails, the value of *c* will accordingly decrease. If $c$ is less than a threshold $f$, which is used to check the correctness of a node, then that node is considered as FAULTY. The detailed detection process is described in **Algorithm 1.**

---

**Algorithm 1** Faulty node detection and node state adjustment algorithm

    Data: **FCS**, confidence variable $c$.
    Result: Node State.
1     Initialize the **FCS** and $c$.
2     **IF** the state of **FCS** is FAULTY or JFAULTY **THEN**
3       Set the state of the Node $i$ as JFAULTY
4     **ELSE**
5     **For each** Node $j$ in **FCS** of Node $i$
6       Node $i$ sends a message to Node $j$
7       **IF** Node $i$ receives no reply from Node $j$ **THEN**
8         Decrease the confidence variable $c$ of Node $j$
9         **IF** the confidence variable $c$ of Node $j$ less than the confidence threshold $f$ **THEN**
10           Set the state of the Node $j$ as FAULTY
11       **ELSE** update the delay and transmission rate of Node $j$

---

*4.2. Network Congestion Detection Method*

**Algorithm 2** is the congestion detection algorithm used in the DMRF protocol. We adopt the similar mechanism as used in [29]. Node buffer utilization and congestion factor are used to predict congestion. The detailed congestion detection description is illustrated in [29]. Congested node will update its own state as **CONG** and feed back this information to its upstream node. If the congested node recovers back to normal state, the information is also fed back to its upstream node. If the **FCS** nodes are all in **CONG** state, then the transmission mode is converted to jumping transmission mode. The detailed congestion detection process is described in **Algorithm 2.**

---

**Algorithm 2** Congestion detection algorithm

    Data: **FCS**, occupy factor of the Node and congestion factor[29]
    Result: Node State.
1     Initialize the **FCS**.
2     Predict the congestion using the method in [29].
3     **IF** Congestion happens Node $i$ **THEN**
4       Set the state of Node $i$ as CONG.
5     **IF** the state of **FCS** is CONG or JCONG **THEN**
6       Set the state of Node $i$ as JCONG, inform to its upstream node.
7     **IF** Node $i$ receives feedback message from Node $j$ **THEN**
8       Set the state of Node $j$ as CONG.
9     **IF** the state of Node $i$ converts to normal **THEN**
10       Set the state of Node $i$ as NORMAL, inform to its upstream node.
11     **IF** Node $i$ receives the updating message **THEN**
12       Node $i$ updates the record in the routing table.

*4.3. VOID Region Detection*

In the DMRF protocol, the VOID region detection method is similar to FTSPEED [5]. If no other nodes exist in the sensing range $r$ of the current node, then the state of the current node is set to VOID, and the information is fed back to its upstream node. When the nodes in **FCS** are all in the VOID state, the state of the current node is also set to VOID. The detailed VOID region detection algorithm is shown in **Algorithm 3**.

| **Algorithm 3** Void region detection algorithm |  |
|---|---|
|  | Data: **FCS** |
|  | Result: Node state |
| 1 | Initialize **FCS**. |
| 2 | **IF** Node *i* has not **FCS THEN** |
| 3 |   Node *i* send feedback message to its upstream Node. |
| 4 | **IF** the states of all Nodes in **FCS** are VOID state **THEN** |
| 5 |   Set the state of Node *i* VOID. |
| 6 | **IF** Node *i* receives feedback message from Node *j* **THEN** |
| 7 |   Set the state of Node *j* VOID. |

*4.4. The Jumping Transmission and Routing Selection*

4.4.1. Jumping transmission

Once node failure, network congestion or VOID region occurs, the transmission mode will switch to jumping transmission mode. The node chooses the next hop node according to jumping probability. Once a jumping transmission fails, each node will re-calculate the corresponding jumping probabilities. Therefore, in the earlier stages of the jumping transmission, transmission failure may occur, but as the transmission proceeds, the transmission failure probability will be reduced greatly. At last each transmission can succeed. The jumping transmission flow chart is shown in Figure 4.

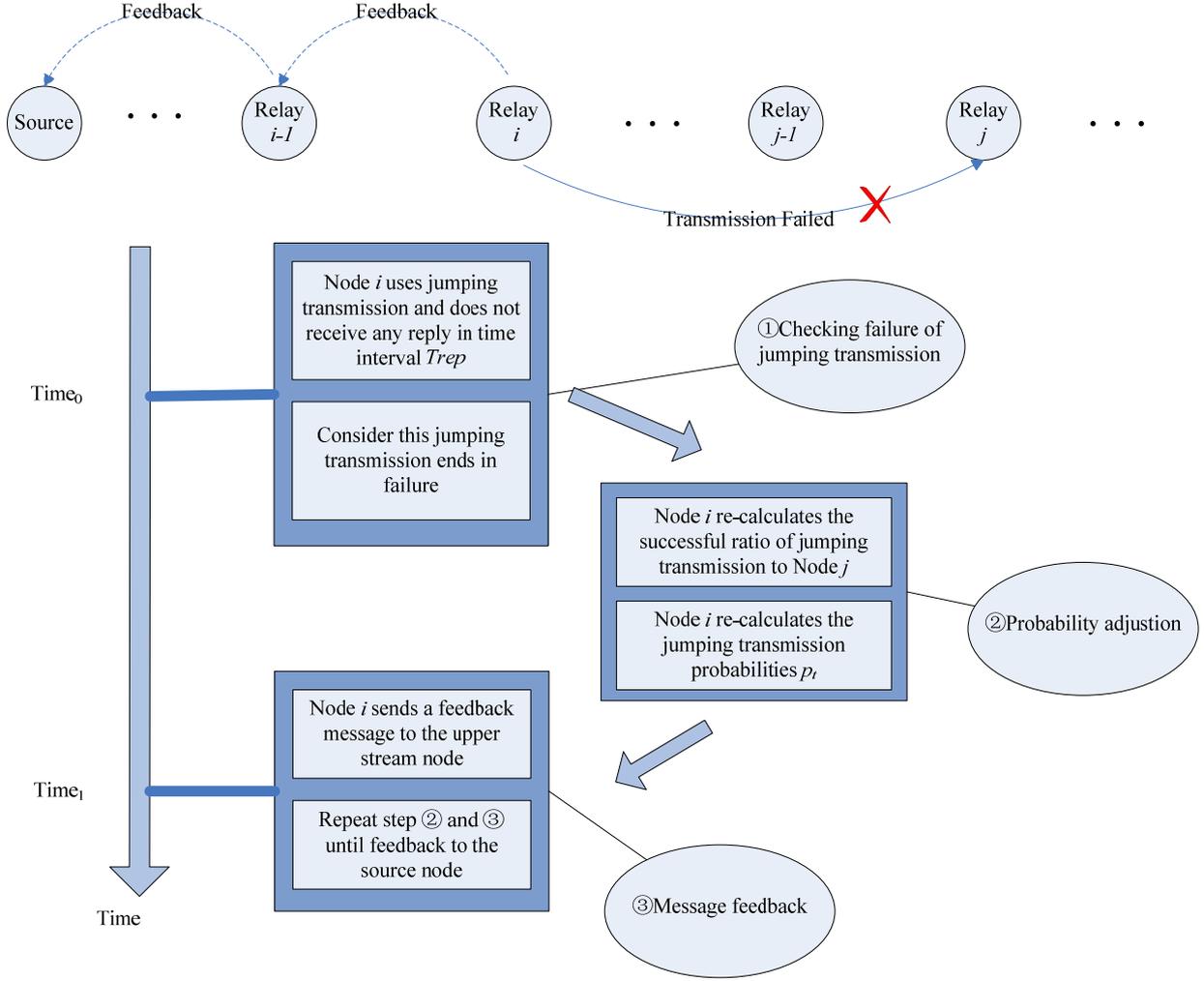

**Figure 4.** Jumping probability adjustment and message feedback.

**(1) Jumping probability adjustment when jumping transmission failure**

If node *i* uses jumping transmission and does not receive any reply from the next hop node in a specified time interval, then the current jumping transmission is considered as failure. $T_t (t = 1, 2, ...m)$ denotes the number of transmissions from the current node to each subsequent node *t* in **FCS**. $S_t$ represents the number of successful transmissions. Accordingly the successful transmission ratio is $Suc_t = S_t / T_t \ (t = 1, 2, 3, ...m)$. Therefore, the jumping probability from the current node to node *t* can be calculated as

$$p_t = \frac{Suc_t}{\sum_{1}^{m} Suc_t} (t = 1, 2, 3, ...m) \tag{1}$$

Once a transmission error occurs, the successful transmission ratio is adjusted as $Suc_t' = (S_t - 1)/T_t (t = 1, 2, 3, ...m)$. Corresponding to the change in the successful transmission ratio, the jumping probability will be recalculated using Equation (1).

Meanwhile, node *i* will send a feedback message to its upstream node. Once receiving the feedback message, the upstream node will also adjust the jumping probability and the feedback message is forwarded hop by hop to the data source node.

(2) Jumping probability adjustment of the upstream node

When the upstream node $i - 1$ receives a feedback message from node *i*, it re-calculates the successful ratio from node $i - 1$ to node *t* as $Suc_t = Suc_t - t$, where $\tau$ is a random factor satisfying $0 \leq \tau \leq Suc_t$. The jumping probability from node $i - 1$ to node *t* can be recalculated using Equation (1).

4.4.2. Path Selection

To guarantee reliable transmission, in the initialization phase of the DMRF protocol, *M* non-overlapping paths from the source to the sink node are established. For each path, each node sends packets to the destination node and calculates the RTT (Round Trip Time) to record the time delay and the distance between the source node and the destination node. By using the heuristic method, *k* paths from the *M* paths are chosen to transmit packets. The paths that are not chosen will be used as alternative paths when necessary. In the process of transmission, when the node receives a data packet, the node will choose the next hop node according to the deadline of the received packet, the transmission delay of the current path, the jumping probability and the congestion level.

Let $T_{minDelay}$ denote the minimum transmission delay from the current node to the destination node. We denote the remaining transmission time factor of the data packet as $\lambda = T_{minDelay}/L$. The choice of the next hop node depends on $\lambda$, $\theta_{low}$ (the lower bound of the remaining transmission time), and $\theta_{high}$ (the upper bound of the remaining transmission time). The algorithm is illustrated in **Algorithm 4**.

---
**Algorithm 4** Selecting the transmission node and transmission mode

Data: **FCS**
Result: Choose the next hop node and the transmitting mode.
1  **IF** the state of node is JFAULTY or VOID or JCONG **THEN**
2      Utilize the jumping mode.
3  **IF** $\lambda \geq \theta_{low}$ **THEN**
4      Select the node with the maximum delay and minimum transmission times. Set the packet state as LOW.
5  **IF** $\theta_{low} > \lambda \geq \theta_{high}$ **THEN**
6      Select the node with the maximum delay and minimum transmission times. Set the packet state as MEDIUM.
7  **IF** $\theta_{high} > \lambda > \theta_{jump}$ **THEN**
8      Select the node with the maximum delay and minimum transmission times. Set the packet state as HIGH.
9  **IF** $\lambda \leq \theta_{jump}$ **THEN**
10     Utilize the jumping mode.

---

The packet states LOW, MEDIUM, and HIGH indicating the three kinds of data packet transmission rates, *i.e.*, low, medium, and high, respectively.

In case $\lambda$ is less than threshold $\theta_{jump}$, the remaining transmission time of the data packet is nearly the same as the transmission delay between the current node and the destination. If the transmission delay in subsequent transmission is too long, the data packet may not be sent to the destination node within the specified time. In this case, the transmission mode is converted to jumping transmission mode. As for other cases (i.e., $\lambda \geq \theta_{low}$, $\theta_{low} > \lambda \geq \theta_{high}$ and $\theta_{high} > \lambda > \theta_{jump}$), the algorithm will adjust the package transmission rate to alleviate the congestion and select reasonably the next hop node with the minimum times of transmission and balance the nodes' energy consumption, thus prolonging the network life time.

Each node calculates $\theta_{low}$ and $\theta_{high}$ according to local information and the data packet information, without need to consider the changes in the entire network. The calculation methods of $\theta_{low}$ and $\theta_{high}$ are given in the Subsection 4.4.3.

### 4.4.3. Calculation of $\theta_{low}$ and $\theta_{high}$

Suppose $P$ nodes are deployed in the path. The transmission time of the node $i$ to the sink node is $T_i$. The delay between node $i$ and node $j$ is $delay_{i,j}$. The remaining transmission time of the data packet in node $i$ is $L_i$.

When the source node collects data packet, a deadline time will be assigned to the data packet. The deadline is larger than the total transmission delay from the source node to the sink node. Therefore, $\lambda \geq 1$. Suppose the communication delay between the node $i$ and its **FCS** in the network conforms to normal distribution, with expected value $\mu$, and node $i$ has $n$ forwarding candidate nodes.

When determining $\theta_{low}$ and $\theta_{high}$, if the current state of data packet is LOW, then the state of the data packet in next hop node can not be HIGH, and vice versa. This requirement can make the node energy consumption smooth because dramatic transmission rate change makes the energy consumption imbalance. In most of the time, $\theta_{low} > \lambda \geq \theta_{high}$.

Therefore, Equation (2) exists.

$$\begin{cases} \dfrac{L_i}{T_i} \geq \theta_{low} \\ \dfrac{L_{i+1}}{T_{i+1}} \geq \theta_{high} \\ \theta_{high} > \dfrac{L_i}{T_i} \geq \theta_{jump} \\ \theta_{low} > \dfrac{L_{i+1}}{T_{i+1}} \geq \theta_{jump} \end{cases} \tag{2}$$

The first two lines in Equation (2) limit that a packet whose status is LOW cannot be converted into the HIGH status in the next transmission process. The next two lines of Equation (2) represent the transmission constraint in which a data packet owns a HIGH status. By analyzing the Equation (2), we work out the following results.

$$\text{Select } \theta_{high} = \theta_{jump} + \frac{\max_{1}^{n}\{delay_{i,n}\}}{T_i}, \theta_{low} = \frac{\theta_{high}}{\omega} + \frac{\mu}{T_i}, \text{ and } \omega = \frac{L_i - \max_{1}^{n}\{delay_{i,i+1}\}}{T_i}.$$

Hence, we can calculate $\theta_{low}$ and $\theta_{high}$ directly using local information, without need to understand the global network topology.

## 5. Feasibility Proof and Performance Analysis of DMRF

In this section, we present the feasibility proof of DMRF and analyze its performance in terms of message complexity and energy consumption complexity.

*5.1. Feasibility Analysis of DMRF Protocol*

Suppose a path contains $m$ nodes, the delay between node $i$ and node $i + 1$ is denoted as $delay_{i,\,i+1}$. The remaining transmission time of each packet when generated from the source is denoted as $L_{source}$.

**Theorem 1:** The DMRF protocol can meet the requirement $\sum_{i=1}^{m-1} delay_{i,i+1} \leq L_{source}$.

*Proof.* We prove **Theorem 1** in two cases: the best case and the worst case.

**The Best case:** In the best case, every hop-by-hop transmission can satisfy $\lambda \geq \theta_{low}$, because the remaining transmission time of the source is larger than the estimated transmission time from the source to the destination. Then we can get $L_{source} \geq T_{source}$. Before the last transmission, it can guarantee that the remaining time $L_{m-1}$ of node $m - 1$ is larger than the estimated transmission time $T_{m-1}$ of node $m - 1$, which is no less than the delay between node $m - 1$ and node $m$. Therefore we can get $L_{m-1} > T_{m-1} \geq delay_{m-1,\,m}$, implying that the transmission can meet the real-time requirement.

**The Worst case:** In the worst case, suppose the remaining transmission time of the data packet is 0 when it is sent to node $k(k < m)$, then $\sum_{1}^{k-1} delay_{i,i+1} \geq L_1$. Each transmission satisfies $\theta_{high} > \lambda > \theta_{jump}$. Consequently $L_i \geq T_i$. Since $T_1 = \sum_{1}^{m-1} delay_{i,i+1}$, we can get $\sum_{1}^{k-1} delay_{i,i+1} > \sum_{1}^{m-1} delay_{i,i+1}$, which means $k > m$. This is in contradiction with $k < m$, thereby the transmission still can meet the real-time performance in the worst case.

Therefore, **Theorem 1** is proved to be correct.

*5.2. Message Complexity Analysis*

**Theorem 2**: The message complexity of DMRF is $O(N)$ in the whole network, where $N$ is the number of nodes in the network.

*Proof.* During the network initialization, each node monitors the notice information of its **FCS** to obtain the information of **FCS**. The monitoring process requires each node to send message once. The complexity is $O(N)$ for this kind of control information. In the feedback procedure, when the states of all nodes in **FCS** are FAULTY or JFAULTY, or congestion and empty area exist, the current node will send feedback message to its upper stream node. In the jumping transmission mode, if the transmission failed, the current node will also send feedback message to its upper stream to inform them to adjust the jumping probabilities. In the worst case, in the feedback procedure within a path with the maximum $S$ hops ($S < N$), each node sends four kinds of feedback messages. Therefore, the control complexity is $O(S)$. It can be seen that the total message complexity of DMRF is $O(N) + O(S) = O(N)$.

Therefore, the **Theorem 2** is proved to be correct.

*5.3. Energy Consumption Complexity Analysis*

**Theorem 3:** $O(\sqrt{N})$ is the energy consumption complexity of the DMRF Protocol.

***Proof.*** Suppose there are $N$ nodes deployed in the network, the average data packet transmission rate of a node is $v$, the maximum remaining transmission time of data packet is $t$, and the average length of one hop is $h$, consequently a node needs to transmit $t/(vh)$ hops. The energy consumption of each hop transmission is $e$. The density of nodes is $d$. Accordingly the average energy consumption complexity of the transmission in a path for SPEED, SPEED-S, SPEED-T, and FTSPEED is

$$O(\frac{t}{vh} \times e) = O(\sqrt{\frac{4N}{\pi d}}) = O(\sqrt{N})$$

As DMRF establishes constant paths in the initial phase, the energy consumption is also $O(1) \times O(\sqrt{N}) = O(\sqrt{N})$.

MMSPEED establishes a tree-like path structure in the initial transmission process. With the distance between the subsequent node and destination node decreasing, the transmission will converge into one path in the latter transmission process. Therefore, its energy consumption complexity is $\frac{1}{2}O(2^{\frac{1}{2}\sqrt{\frac{4N}{\pi d}}}) + \frac{1}{2}O(\sqrt{N}) = O(2^{\sqrt{N}})$, which is higher than DMRF, SPEED, and FTSPEED.

*5.4. Time Complexity of Faulty Nodes, Congested Nodes and VOID Region Detection*

**Theorem 4**: $O(N)$ is the time complexity of detecting faulty nodes, congested nodes and VOID region.
***Proof.*** Suppose $N$ nodes are deployed in the network uniformly. In the initialization phase, each node will establish its **FCS** and initialize the status of the nods in **FCS**. Since all nodes are uniformly distributed, the average size of each **FCS** can be considered as a constant $f_{cs}$. Therefore the complexity of detecting faulty nodes, congestion nodes and VOID region in the initial phase is $O(f_{cs} \times N) = O(N)$.

When faulty node, congested nodes or VOID region occur in the network, the upper stream nodes will receive the feedback message, which means the detecting complexity is the identical with the complexity of the control messages. Therefore, the time complexity of detecting faulty nodes, congested nodes and VOID region is $O(N)$.

## 6. Performance Evaluation

In this section, we present the results of several simulations to evaluate the performance of the DMRF protocol. We use the JProwler simulation platform [30], which is an event-driven WSN-based simulation platform. Consider a typical wireless sensor network in a real-time transmission scenario. Nodes conform to uniform or random distribution. During the experiments, the source node periodically sends data to the sink node. A certain amount of failure nodes are deployed. During the transmission, the buffer occupancy of partial nodes increases and network congestion occurs, causing some data packets to be discarded due to deadline miss. All the parameters used in the simulations are shown in Table 2.

**Table 2.** The setting of parameters used in the simulation.

| Routing Protocol | SPEED, SPEED-T, SPEED-S, MMSPEED, FTSPEED, DMRF |
|---|---|
| MAC Layer | 802.11 |
| MAC Protocol | CSMA/CA |
| Bandwidth | 200 Kb/s |
| Buffer Size | 100 Bytes |
| Data packet Size | 32 Bytes |
| Region Size | (20 m, 20 m) |
| Node Number | 400 |
| Node Distribution | Random and Uniform |
| Maximum Transmission Distance | 30 m |

*6.1. Times of Successful Transmission with Different Ratios of Faulty Nodes*

We randomly inject a certain ratio of faulty nodes in the network, the initial buffer size is set as 0, *i.e.*, there is no congestion. 100 data packets are sent from the source node to the destination node. Times of successful transmission are recorded to reflect the impact of faulty nodes on data transmission. The comparison between SPEED, SPEED-S, SPEED-T, MMSPEED, FTSPEED and DMRF is shown in Figure 5. Because the VOID region in the network only yields to faulty nodes or the change of the nodes distribution (random distribution), this experiment can also reflect the capacities to avoid the influence of the VOID region among these six protocols.

In Figure 5, we can see that with the ratio of faulty nodes increasing, the times of successful transmission of most methods decline, whatever distribution the nodes conform to. In the random distribution case, an even more dramatic decline occurs, which is mainly due to the uneven distribution of nodes causing larger empty regions and leading to the transmission failure. DMRF establishes multiple non-overlapping paths, which can reduce the impact of node failure. In the interval [0.2,0.4], the number of successful transmissions increases for the DMRF protocol. The jumping transmission mode adopted in DMRF reduces the impact of node failure greatly. Moreover, the jumping probability will be adjusted according to the transmission result (success or failure). With the transmission going on, the times of successful transmission will increase. As regarding SPEED-S and SPEED-T, only the selection method of the next hop node is different from SPEED. Therefore the overall trend is similar to the results of SPEED. In MMSPEED, the transmission failure ratio is higher because nodes in the initial transmission cannot meet the reachability requirements. According to the results, we can see that DMRF exhibits higher capacity in avoiding VOID region.

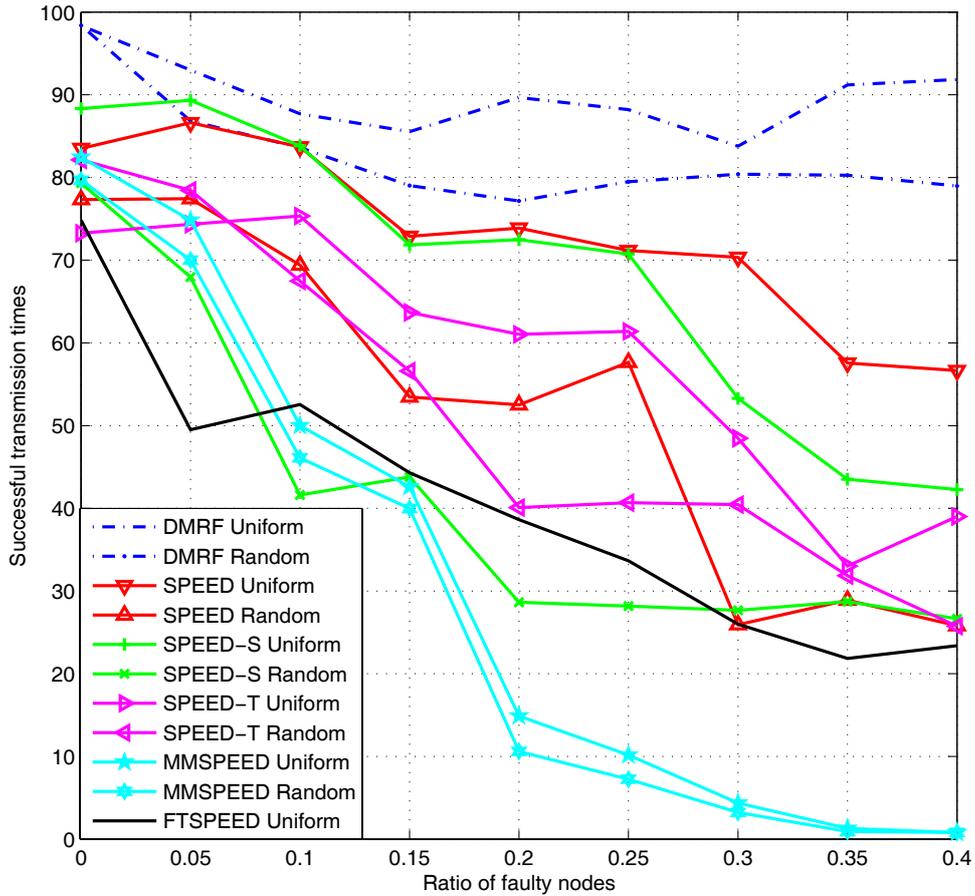

**Figure 5.** The impact of faulty nodes on network transmission.

*6.2. Successful Transmission Ratio under Congestion Condition*

As can be seen from Figure 6, with the initialization ratio of the buffer increasing, the times of successful transmission reduce gradually. In this experiment, we just omit the effect of faulty nodes, so the probability of the faulty node is set 0. Therefore the VOID region can be treated as congested nodes or the change of the nodes distribution (random distribution).

The decrease of all the methods except MMSPEED is flat. When the next hop is a congested node, FTSPEED applies a bypass way to relieve the effect of the congestion node. But this method can only achieve the local solution of reducing the effect of congestion nodes. For SPEED, it simply discards the package when congestion occurs. As depicted in Figure 6, the times of successful transmission of DMRF is obviously higher than that of the SPEED. The main reason is that DMRF can jump over the congestion node at a degree of probability. But the successful ratio is not always 100%, since the packet can be sent to another congestion node. As the data packet is sent, the jumping probabilities are adjusted, which increases the ratio of successful jumping transmission.

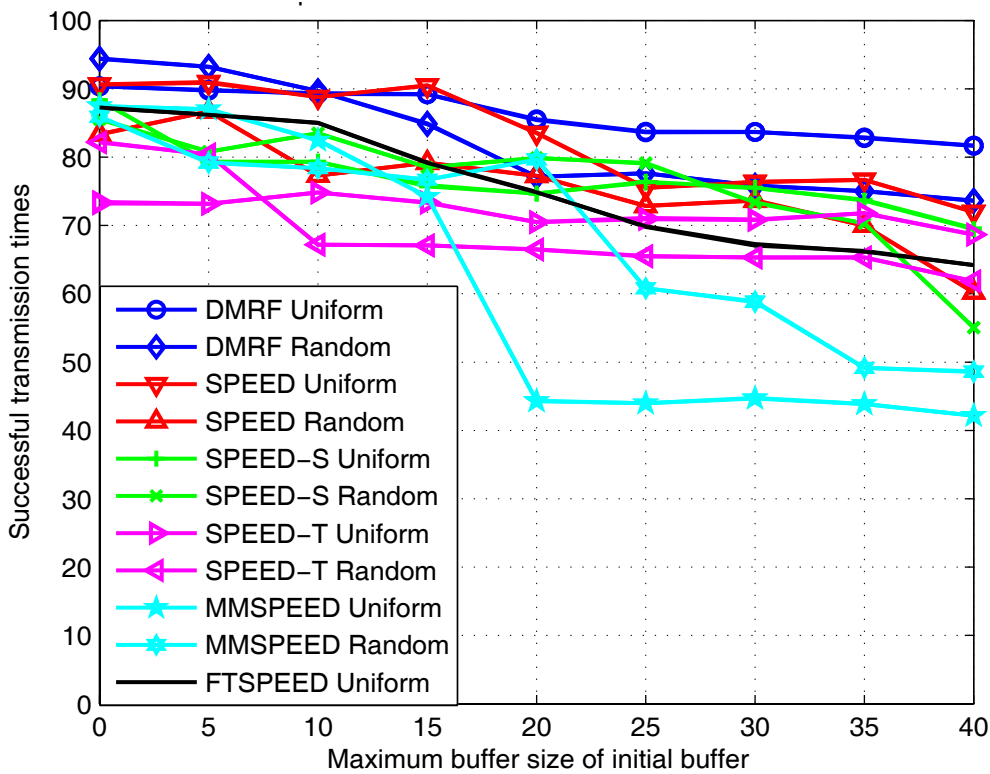

**Figure 6.** The impact of network congestion on transmission.

*6.3. The Effect of Network Topology on DMRF*

In Figure 5, there is not much change in successful transmission times. This shows that no matter whether the topology of the sensor networks is uniform or not, the reachability of DMRF does not obviously change. The jumping transmission mode utilized in DMRF can confront the effect of network topology on transmission performances. In Figure 6, it also shows the network topology does not affect the successful transmission times of DMRF. Therefore, DMRF can resist the impact of network topology effectively.

*6.4. Successful Transmission Ratio in Case of VOID Region*

We compare SPEED, MMSPEED, FTSPEED with DMRF in terms of the impact of void region on the successful transmission ratio in the case of uniform distribution. The simulation results are shown in Figure 7.

Since DMRF dynamically adjusts the jumping probability, the packet transmission may fail because of void region in the original transmission. But in later transmissions, the probability of the successful transmission increases greatly. In Figure 7, when the radius of void region increases to seven, only FTSPEED and DMRF can still transmit data. DMRF can achieve the successful transmission ratio up to 92%. When the radius is eight, DMRF directly transmits the data packet from the source to the destination node. Although the energy consumption of this kind of transmission is high, it is the only way to achieve successful transmission.

*6.5. Average Transmission Delay in Case of Different VOID Region Radius*

The average delay of data transmission with respect to void area radius change is shown in Figure 8. In Figure 7 and 8, we can see when the radius exceeds six under SPEED, SPEED-T, SPEED-S, and MMSPEED, the network transmission delay significantly increases while the times of successful transmission significantly decreases, and serious congestion occurs, which causes a large number of data packets to be discarded. In contrast, DMRF can switch to jumping transmission mode and jump the void region or congested areas, hereby the packet transmission delay has no significant change.

**Figure 7.** Influence of Void region on transmission.

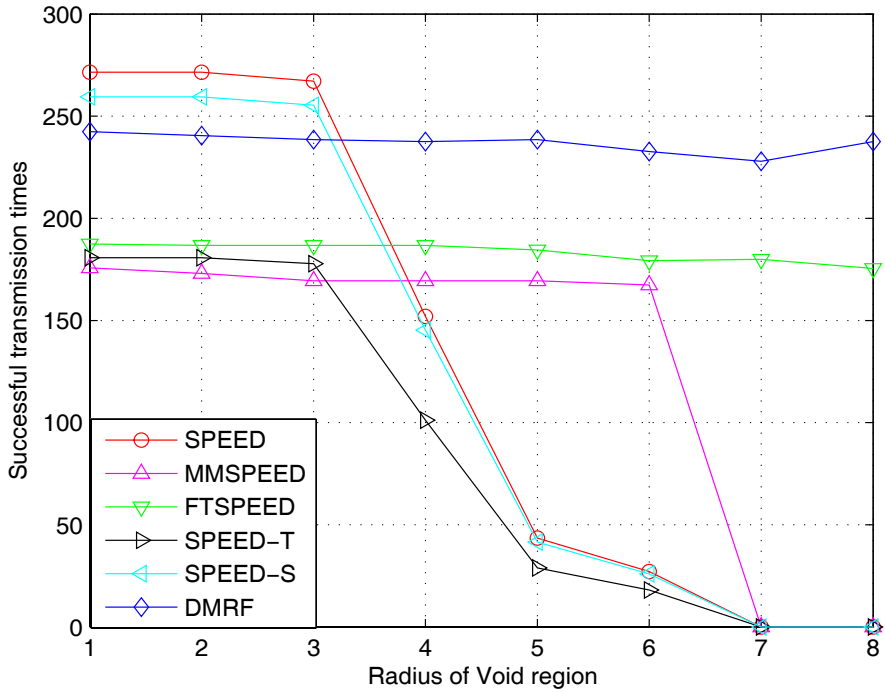

*6.6. The Number of Control Packets in Case of Different VOID Region Radius*

In DMRF, when node failure, void area or network congestion occurs, the current node feeds back the information to its upstream node immediately. The upstream node then adjusts its jumping probability. In Figure 9, when the radius of the void region is in (0, 5], the number of control packets increases. When the radius is above five, the number of control packets declines due to the decrease of the number of working nodes. It can be seen from Figure 9, that DMRF uses more control packets than other methods when the radius of the void region is in [1, 4), because DMRF uses three kinds of feedback packets. When the radius is more than four, network congestion occurs, and SPEED, SPEED-T, SPEED-S, and MMSPEED begin to discard packets. Therefore, the number of control packets increases dramatically. In total, the number of control packets of DMRF is related to the number of nodes working in the area, instead of the global network topology. Therefore, the number of control packets is relatively stable. The number of control packets of DMRF is slightly higher than that of FTSPEED, but far less than those of SPEED, SPEED-T, and SPEED-S. That is, DMRF still has certain superiority in reducing the number of control packets.

**Figure 8.** Relationship between transmission delay and Void region.

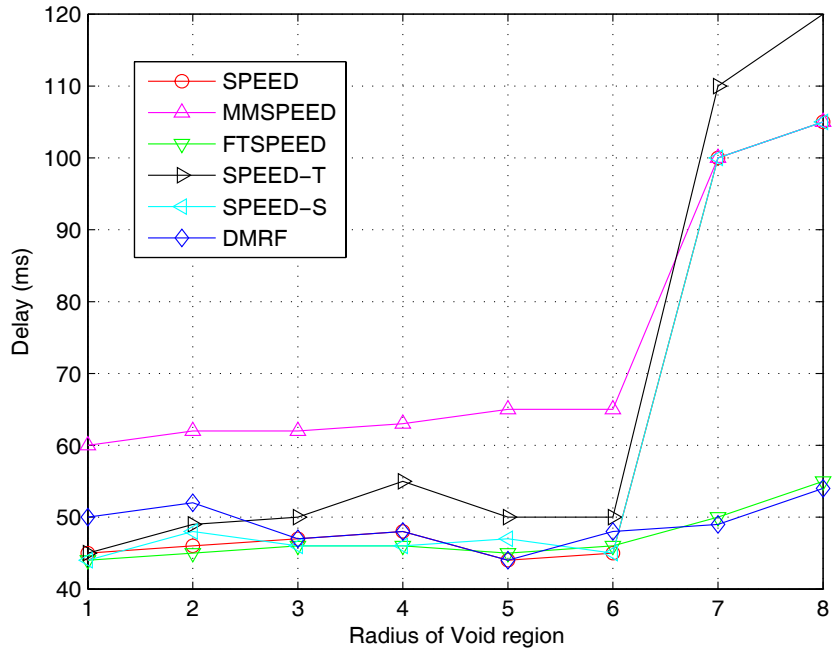

**Figure 9.** Relationship between Void region and control packets.

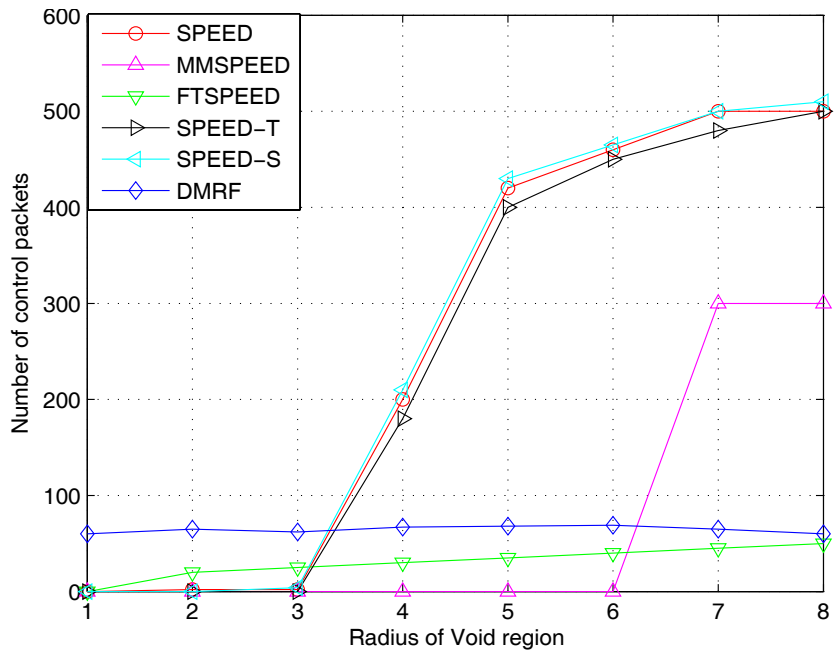

## 7. Conclusions and Future Work

In this paper a dynamical jumping real-time fault-tolerant routing protocol (DMRF) has been proposed. DMRF works in two data transmission modes: hop-by-hop mode and jumping transmission mode. Once node failure, network congestion or empty region occurs, the transmission mode will switch to jumping mode, which can reduce the transmission time delay, ensuring the data packets to be sent to the sink node within the specified time limit. We theoretically prove that DMRF can meet real-time and fault tolerance requirements. The performance of DMRF is evaluated by extensive simulation experiments. Simulation results show that DMRF can not only efficiently reduce the effects of failure nodes, congestion and void area, but also increase the ratio of successful transmission, lower the transmission delay, and reduce the number of the control packets. The primary contributions of this paper are summarized as follows.

(1) The jumping transmission mode is explored to guarantee real-time and fault-tolerant characteristics.
(2) Feedback mechanism is used to enhance the successful transmission ratio.
(3) The average energy cost of each node in the network is balanced and the life time of the whole network is prolonged by the selection method of next hop in which the node in FCS with the minimum times of transmission is selected as the next hop.
(4) The feasibility proof and performance analysis are presented to testify the superiority of DMRF.

As part of our future work, we would like to extend DMRF for dynamic WSN environments, where fault tolerance characteristic is one of the most desirable features that should be provided. It is also valuable to revise the DMRF protocol for cluster-based real-time routing in order to achieve much lower energy consumption.

## Acknowledgements

This work was partially supported by the National Natural Science Foundation of China under Grant No. 60703101 and No. 60903153.